\documentclass[prb,aps,twocolumn,floatfix]{revtex4}
\usepackage{amsmath}
\usepackage{amssymb}
\usepackage{amsthm}
\usepackage{epsfig}
\usepackage{tikz}
\usepackage{enumitem}
\usepackage{etoolbox}
\usepackage{xcolor}

\def\up{\uparrow}
\def\pd{ {\phantom\dagger} }
\def\down{\downarrow}
\def\nn{\nonumber\\}
\def\fa#1#2{{a_#1(#2)}}
\def\l{\Lambda}

\newcommand{\be}{\begin{equation}}
\newcommand{\ee}{\end{equation}}
\newcommand{\beA}{\begin{equation}\begin{aligned}}
\newcommand{\eeA}{\end{aligned}\end{equation}}
\newcommand{\bem}{\begin{multline}}
\newcommand{\eem}{\end{multline}}
\newcommand{\bea}{\begin{eqnarray}}
\newcommand{\eea}{\end{eqnarray}}
\theoremstyle{definition}

\theoremstyle{remark}

\def\fr#1{(\ref{#1})}

\def\ontop#1#2{\genfrac{}{}{0pt}{}{#1}{#2}}

\begin{document}
%%%%%%%%%%%%%%%%%%%%%%%%%%%%%%%%%%%%%%%%%%%%%%%%%%%%%%%%%%%%%%%%%%%%%
\title{Quantum-disentangled liquid in the half-filled Hubbard model} 
%%%%%%%%%%%%%%%%%%%%%%%%%%%%%%%%%%%%%%%%%%%%%%%%%%%%%%%%%%%%%%%%%%%%%
\author{Thomas Veness}
\affiliation{The Rudolf Peierls Centre for Theoretical Physics, University of
Oxford, Oxford OX1 3NP, UK}

\author{Fabian H.~L.~Essler}
\affiliation{The Rudolf Peierls Centre for Theoretical Physics,
  University of Oxford, Oxford OX1 3NP, UK}

\author{Matthew P.A. Fisher}
\affiliation{Department of Physics, University of California, Santa
Barbara, California 93106, USA}

\begin{abstract}
We investigate the existence of quantum disentangled liquid (QDL)
states in the half-filled Hubbard model on bipartite
lattices. In the one dimensional case we employ a combination of
integrability and strong coupling expansion methods to argue that
there are indeed finite energy-density eigenstates that exhibit QDL
behaviour in the sense of J. Stat. Mech. P10010 (2014).
%In particular we show that measuring the spin degrees of
%freedom in these states leaves the system in an area-law entangled
%state at all orders in the expansion, if it converges. 
The states exhibiting the QDL property are atypical in the sense that
while their entropy density is non-zero, it is smaller than that of
thermal states at the same energy density. We argue that for $U\gg t$
these latter  thermal states exhibit a weaker form of the QDL
property, which carries over to the higher dimensional case. 

\end{abstract}

\maketitle
\section{Introduction}
%\emph{Introduction.} 
The last decade has witnessed significant
advances in our understanding of relaxation in isolated many-particle
quantum systems. Many of these were driven by advances in ultra-cold
atom experiments, which have made it possible to study the real-time
dynamics of almost isolated systems in exquisite detail \cite{GM:col_rev02,kww-06,HL:Bose07,hacker10,tetal-11,getal-11,cetal-12,langen13,MM:Ising13,zoran1,MBLex1}.
At present there are three established paradigms for the relaxational
behaviour in such systems. According to the
eigenstate thermalization hypothesis \cite{Deutsch91,Sred1,Sred2,ETH}
``generic'' systems relax towards thermal equilibrium
distributions. In this case the only information about the
initial state that is retained at late times is its energy
density. Quantum integrable models cannot thermalize in this way as
they exhibit additional conservation laws, but they relax to
generalized Gibbs ensembles instead
\cite{GGE,EFreview,Cazalilla06,cc-07,CEF,GGE_int}. Finally, disordered 
many-body localized systems fail to
thermalize as well\cite{MBL1,MBL2,MBL3,MBL4,MBL5,MBL6}. This is
related to the existence of conservation laws\cite{MBL6,MBL7,MBL8},
although in contrast to integrable models no fine-tuning of the
Hamiltonian is required. 

Recently it has been proposed that certain
isolated quantum systems may exhibit behaviour that is not
captured by these three known paradigms\cite{qdl}. The resulting
state of matter has been termed a ``quantum disentangled liquid'' (QDL).
A characteristic feature of such systems is that they comprise both
``heavy'' and ``light'' degrees of freedom, which will be made 
more precise below. The basic premise of the QDL
concept is that while the heavy degrees of freedom are fully
thermalized, the light degrees of freedom, which are enslaved to the
heavy degrees of freedom, are not independently thermalized. 
A convenient diagnostic for such a state of matter is the bipartite
entanglement entropy (EE) after a projective measurement of the heavy particles.
Specifically, we define a QDL to be a state in which the entanglement entropy
reduces from a volume-law to an area-law upon a projective measurement
of the heavy particles.
The possibility of realizing a QDL in the one dimensional Hubbard
model was subsequently investigated by exact diagonalization of small
systems in Ref.~\onlinecite{hubb1}. Given the limitations on
accessible system sizes it is difficult to draw definite conclusions
from these results. Motivated by these studies we explore the
possibility of realizing a QDL in the half-filled Hubbard model on
bipartite lattices by analytic means. We employ methods of
integrability as well as strong-coupling expansion
techniques~\cite{takahashi,MacDonaldGirvin88,stein,sasha} to analyze
properties of finite energy-density eigenstates of the Hubbard
Hamiltonian 
\bea
H&=&-t\sum_{\langle j,k\rangle}\sum_{\sigma=\up,\down}
c^\dagger_{j,\sigma}c^\pd_{k,\sigma}%+{\rm h.c.}
+c^\dagger_{k,\sigma}c^\pd_{j,\sigma}\nn
&+&U\sum_{j}
\big(n_{j,\uparrow}-\frac{1}{2}\big)
\big(n_{j,\downarrow}-\frac{1}{2}\big).
\label{HHubb}
\eea
Here $n_{j,\sigma}=c^\dagger_{j,\sigma}c^\pd_{j,\sigma}$ and $\langle
j,k\rangle$ denote nearest neighbour links that connect only sites from
different sub-lattices $A$ and $B$. At half-filling (one electron per
site) \fr{HHubb} is invariant under the $\eta$-pairing SU(2) symmetry
with generators\cite{etapairing} 
\bea
\eta^z&=&\frac{1}{2}\sum_j \left(n_{j,\up}+n_{j,\down}-1\right)\ ,\nn
\eta^\dagger&=&\sum_{j\in A}c^\dagger_{j,\up}c^\dagger_{j,\down}
-\sum_{k\in B}c^\dagger_{k,\up}c^\dagger_{k,\down}\equiv \left(\eta\right)^\dagger.
\eea
In the strong interaction limit of the Hubbard model ($U\gg t$), the 
bandwidths of the collective spin and charge degrees of freedom are
proportional to $t^2/U$ and $t$ respectively. In this regime the spin
(charge) degrees of freedom therefore correspond to ``heavy'' (``light'')
particles. 

The outline of our paper is as follows. In section
\ref{sec:integrability} we employ methods of integrability to show
that it is possible to construct certain (``Heisenberg sector'')
macrostates at finite energy densities for which the charge degrees of
freedom do not contribute to the volume term in the bipartite EE. This
strongly suggests that these states have the QDL property. In section
\ref{sec:strongcoupling} we then turn to strong-coupling expansion
methods in order to examine the QDL diagnostic proposed in
Ref.~\onlinecite{qdl}. In the one dimensional case this analysis shows
that in the framework of a $t/U$-expansion a projective measurement of
the spin degrees of freedom in a Heisenberg sector state indeed leaves
the system is a state that is only area-law entangled. Our
strong-coupling analysis further suggests the existence of a weaker version
of the QDL property, where a projective measurement of the heavy
degrees of freedom leaves the system in a state characterized by a
volume-law entanglement entropy, but with a pre-factor that is
exponentially small in $U/t$.
Section \ref{sec:conc} contains a summary and discussion of our
results. Various technical aspects of our calculations are presented
in three appendices.

%%%%%%%%%%%%%%%%%%%%%%%%
\section{Integrability}
\label{sec:integrability}
%%%%%%%%%%%%%%%%%%%%%%%%
The one dimensional Hubbard model is exactly solvable by the Bethe
ansatz method \cite{book}. This provides a description of energy
eigenstates as well as the  corresponding macro states in the
thermodynamic limit. We now use this framework to identify a class of
macro states that exhibits a property that is characteristic of a
QDL. 
%%%%%%%%%%%%%%%%%%%%%%%%%%%%%%%%%%%%%%%%%%%%%%%%%%%%%%%%%%%%
%\section{Bethe Ansatz form of Heisenberg sector states}
%\label{app:Hubbard}
%%%%%%%%%%%%%%%%%%%%%%%%%%%%%%%%%%%%%%%%%%%%%%%%%%%%%%%%%%%%
In the framework of the string hypothesis general macro states in the
one dimensional Hubbard model are described by sets of particle and
hole densities
$\{\rho^{p}(k), \rho^{h}(k),
\sigma^{p}_n(\Lambda), \sigma^{h}_n(\Lambda),
{\sigma'_n}^{p}(\Lambda), {\sigma'_n}^{h}(\Lambda)
|n\in\mathbb{N}\}$.
These are analogs of the well-known particle and hole densities used to
characterize macro states in the ideal Fermi gas. The main
complication in integrable models is that there are in general
(infinitely) many different species of excitations, which interact
with one another. In the case at hand the $\rho^{p/h}$ describe
pure charge excitations, the $\sigma^{p/h}_n$ correspond to
elementary spin excitations as well as their bound states and
${\sigma'_n}^{p/h}$ represent bound states between spin and charge
degrees of freedom. As a consequence of the interacting nature of the
Hubbard chain the root densities are related in a non-trivial manner:
they are subject to the thermodynamic limit of the Bethe Ansatz equations\cite{book}
\begin{widetext}
\index{root density}
\begin{eqnarray}
&&\rho^p(k)+\rho^h(k)=\frac{1}{2\pi}+\cos
k\sum_{n=1}^\infty\int_{-\infty}^\infty d\Lambda\ \fa{n}{\l-\sin k}
\left[{\sigma^\prime_n}^p(\Lambda)+\sigma_n^p(\Lambda)\right]\ ,\nn
&&\sigma_n^h(\Lambda)=-\sum_{m=1}^\infty
\int_{-\infty}^\infty d\Lambda' A_{nm}(\Lambda-\Lambda')\ \sigma_m^p(\Lambda')
% A_{nm}*\sigma_m^p\bigg|_\Lambda
+\int_{-\pi}^\pi dk\ \fa{n}{\sin k-\Lambda}\ \rho^p(k)\ ,\nn
&&{\sigma^\prime_n}^h(\Lambda)=\frac{1}{\pi}{\rm
Re}\frac{1}{\sqrt{1-(\Lambda -inu)^2}}
-\sum_{m=1}^\infty 
\int_{-\infty}^\infty d\Lambda' A_{nm}(\Lambda-\Lambda')\ {\sigma'_m}^p(\Lambda')
%A_{nm}*{\sigma^\prime_m}^p\bigg|_\Lambda
-\int_{-\pi}^\pi dk\ \fa{n}{\sin k-\Lambda}\ \rho^p(k)\ .
\label{densities}
\end{eqnarray}
\end{widetext}
Here $u=U/4t$ and
\bea
a_n(x)&=&\frac{1}{2\pi}\frac{2nu}{(nu)^2+x^2}\ ,\nn
\label{an}
A_{nm}(x)&=&\delta(x)+ (1-\delta_{m,n})a_{|n-m|}(x)+2a_{|n-m|+2}(x)\nn
&+&\dots+2a_{|n+m|-2}(x)+a_{n+m}(x).
\label{Anm}
\eea
The energy and thermodynamic entropy per site are then given by
\bea
e&=&u+
\int_{-\pi}^\pi dk\left[-2\cos k -\mu-2u\right]\rho^p(k)\ \nn
&+& 4 \sum_{n=1}^\infty \int d\Lambda\,{\sigma^\prime_n}^p(\Lambda)\,{\rm Re}\sqrt{1-(\Lambda+inu)^2} 
,\nn
s&=&\int_{-\pi}^\pi dk\ {\cal S}\left[\rho^p(k),\rho^h(k)\right]\nn
&+&\sum_{n=1}^\infty\int_{-\infty}^\infty d\Lambda\
{\cal S}\left[{\sigma^\prime_n}^p(\Lambda),{\sigma_n^\prime}^h(\Lambda)\right]\nn
&+&\sum_{n=1}^\infty\int_{-\infty}^\infty d\Lambda\
{\cal S}\left[{\sigma_n}^p(\Lambda),{\sigma_n}^h(\Lambda)\right],
\label{entropy}
\eea
where we have defined
\bea
{\cal S}[f,g]&=&
\big[f(x)+g(x)\big]\ln\big(f(x)+g(x)\big)\nn
&-&f(x)\ln\big(f(x)\big)
-g(x)\ln\big(g(x)\big)\ .
\eea
The ground state of the half-filled Hubbard model in zero magnetic
field is obtained by choosing
\be
\rho^h(k)=0=\sigma^h_1(\Lambda)\ ,\qquad
{\sigma'}_n^p(\Lambda)=0=
{\sigma}_{n\geq 2}^p(\Lambda)\ .
\ee
%%%%%%%%%%%%%%%%%%%%%%%%%%%%%%%%%%%%%%
\subsection{The ``Heisenberg sector''}
%%%%%%%%%%%%%%%%%%%%%%%%%%%%%%%%%%%%%%
Motivated by Ref.~\onlinecite{hubb1} we now consider a particular
class of macro states, which we call \emph{Heisenberg sector states}. 
The principle underlying their construction is to ``freeze'' the charge
degrees of freedom in its ground state configuration while imposing a
finite energy density in the spin sector. This is achieved by
requiring
\be
\rho^h(k)=0={\sigma'}_n^p(\Lambda)\ ,\ n=1,2,\dots
\label{HeisenbergBA}
\ee
Condition \fr{HeisenbergBA} corresponds to the absence of bound states
between spin and charge degrees of freedom (i.e. no $k$-$\Lambda$
strings) and a completely filled ``Fermi sea'' of elementary charge
degrees of freedom. Importantly the thermodynamic
entropy per site of Heisenberg sector states depends only on the spin
sector 
\be
s=\sum_{n=1}^\infty\int_{-\infty}^\infty d\Lambda\
{\cal S}\left[{\sigma_n}^p(\Lambda),{\sigma_n}^h(\Lambda)\right].
\label{SH}
\ee
As shown in Appendix~\ref{app:counting} the total number of 
Heisenberg sector states $N_{\rm HS}$ fulfils
\be
\lim_{L\to\infty}\frac{\ln\big(N_{\rm HS}\big)}{L}=\ln(2).
\label{HScount}
\ee
The result \fr{HScount} suggests that for large, finite $L$ there are
approximately $2^L$ Heisenberg sector states. 
%%%%%%%%%%%%%%%%%%%%%%%%%%%%%%%%%%%%%%%%%%%%%%%%%%%%%%%%%%%%%%%%
\subsection{Entanglement entropy of Heisenberg sector states}
%%%%%%%%%%%%%%%%%%%%%%%%%%%%%%%%%%%%%%%%%%%%%%%%%%%%%%%%%%%%%%%
We now make use of the relation between the volume term in the EE and
the thermodynamic entropy density for eigenstates of short-ranged
Hamiltonians. Consider a finite energy density eigenstate
$|\Psi\rangle$ and a large subsystem $A$ of size $|A|$. Then the
volume term in the EE of the state $|\Psi\rangle$ is given by
\be
S_{\rm vN,A}=s|A|+o(|A|)\ .
\label{SA}
\ee
where $s$ is the thermodynamic entropy density of the macro state
associated with $|\Psi\rangle$.

Combining \fr{SH} with
\fr{SA}, we conclude that \emph{for Heisenberg sector states the
  volume term in the entanglement entropy is entirely due to the spin
  degrees of freedom, and the charge degrees of freedom do not
  contribute}. This is very much in line with what one would expect
for a QDL state.
%%%%%%%%%%%%%%%%%%%%%%%%%%%%%%%%%%%%%%%%%%%%%%%%%%%%%%%%%%%%%%%%
\subsection{Thermal states in the large-$U$ limit}
%%%%%%%%%%%%%%%%%%%%%%%%%%%%%%%%%%%%%%%%%%%%%%%%%%%%%%%%%%%%%%%%
It is very instructive to contrast the entanglement properties of
Heisenberg sector states to those of typical states.
% at low energy densities. 
The maximum entropy states at a given energy density are
thermal and can be constructed by the Thermodynamic Bethe Ansatz (TBA).
This provides a system of coupled non-linear integral equations 
for the ratios of the hole and particle
densities\cite{book,takahashiTBA1} 
\be
\zeta(k)=\frac{\rho^h(k)}{\rho^p(k)}\ ,\
\eta_n(\Lambda)=\frac{\sigma_n^h(\Lambda)}{\sigma^p_n(\Lambda)}\ ,\
\eta'_n(\Lambda)=\frac{{\sigma'_n}^h(\Lambda)}{{\sigma'_n}^p(\Lambda)}\ .
\label{dressedE}
\ee
For the sake of completeness we present the TBA
equations in Appendix~\ref{app:TBAeqns}. While in general the TBA
equations can only be solved numerically, in the limit of strong
interactions analytic results can be obtained
\cite{takahashiTBA2,Ha,EEG}. For simplicity we focus on the
``spin-disordered regime'' 
\be
\frac{4t^2}{U}\ll T\ll U\ .
\ee
This regime corresponds to temperatures that are small compared to
the Mott gap, but large compared to the exchange energy. Here one
has\cite{EEG} 
\be
\rho^h(k)={\cal O}\big(e^{-u/T}\big)\ ,\quad
{\sigma'}^{p,h}_n(\Lambda)={\cal O}\big(e^{-u/T}\big)\ ,
\ee
where we have set $t=1$ as our energy scale.
Substituting this into the general expression \fr{entropy} for the
thermodynamic entropy density we obtain
\be
s=\sum_{n=1}^\infty\int_{-\infty}^\infty d\Lambda\
{\cal S}\left[{\sigma_n}^p(\Lambda),{\sigma_n}^h(\Lambda)\right]
+{\cal O}\Big(\frac{u}{T}e^{-u/T}\Big).
\label{ST}
\ee
Finally, using the relation between thermodynamic and EE
\fr{SA} we conclude that for thermal states in the
spin-disordered regime the contribution of the charge degrees of
freedom contribute to the volume term is
%\be
%S_{\rm vN,A}\Bigg|_{\rm charge}={\cal O}\Big(\frac{u}{T}e^{-u/T}\Big)|A|.
%\ee
\bea
S_{\rm vN,A}&=&\big(s_{\rm spin}+s_{\rm
  charge}\big)|A|+o\big(|A|\big)\ ,\nn
s_{\rm spin}&=&{\cal O}(1)\ ,\quad
s_{\rm charge}={\cal O}\Big(\frac{u}{T}e^{-u/T}\Big)\ .
\label{weak}
\eea
Here $s_{\rm charge}$ includes the contributions from pure charge
degrees of freedom as well as bound states of spin and charge.
Importantly, unlike Heisenberg sector states, typical states have a
contribution from the charge degrees of freedom to the volume
term. However, this contribution is exponentially small in $u/T$ and
therefore only visible for extremely large subsystems. While we have
focussed on the spin-disordered regime, the behaviour \fr{weak}
extends to thermal states for all $0<T\ll U$.
%%%%%%%%%%%%%%%%%%%%%%%%%%
\section{$t/U$-Expansion} 
\label{sec:strongcoupling}
%%%%%%%%%%%%%%%%%%%%%%%%%%
%\emph{$t/U$-Expansion.} 
The analysis presented above provides a strong indication that the
Heisenberg sector states realize the QDL concept. However, the exact
solution does not presently allow one to examine the QDL diagnostic
proposed in Ref.~\onlinecite{qdl}, which requires the calculation of
the EE after a partial measurement. We therefore now turn to a
complementary approach, namely a strong-coupling expansion in powers
of $t/U$. This is most conveniently implemented by following
Ref.~\onlinecite{MacDonaldGirvin88}. At a given site $j$ there are 
four possible states $|0\rangle_j$, $|{\up}\rangle_j =
c_{j,\up}^\dagger |0\rangle_j$, $|{\down}\rangle_j =
c_{j,\down}^\dagger |0\rangle_j$ and $|2\rangle_j = c_{j,\up}^\dagger
c_{j,\down}^\dagger|0\rangle_j$. Defining Hubbard operators by
$X_{j}^{ab} = |a\rangle_j { }_j\langle b|$, the Hamiltonian
\fr{HHubb} can be expressed in the form 
\be
H = UD + t\left(   T_0 +  T_1 +  T_{-1}\right),
\ee
where $D=\frac{1}{4} \sum_jX^{22}_j + X^{00}_j - X^{\up\up}_j 
- X^{\down\down}_j$ and $T_{a}=\sum_jT_{a,j}$ are correlated
hopping terms that change the number of doubly occupied sites by $a$
\beA
T_{0,j} &= - \sum_{\sigma}  \big(
X^{2\sigma}_j X^{\sigma2}_{j+1}
+ X_j^{\sigma0} X_{j+1}^{0\sigma}
+{\rm h.c.}\big) ,\nn
T_{1,j} &= T_{-1,j}^\dagger=- \sum_{\sigma} 
\sigma\left[ X_j^{2\bar{\sigma}} X_{j+1}^{0\sigma} 
+ X_j^{0\bar{\sigma}} X_{j+1}^{2\sigma} \right]\ .
\eeA
The $t/U$-expansion is conveniently cast in the form of a unitary transformation
\cite{MacDonaldGirvin88}
\be
H' = e^{iS} H e^{-iS} = H +[iS,H] + \frac{1}{2} [iS,[iS,H]] + \dots,
\ee
where the generator $iS$ is chosen as a power series in $t/U$ 
$iS = \sum_{n=1}^\infty\left(\frac{t}{U}\right)^n iS^{[n]}$.
The operators $S^{[1]},\dots, S^{[k]}$ can be chosen such that the
first $k+1$ terms in the $t/U$-expansion of $H'$ will not change the
number of doubly occupied sites. It follows from
Ref.~\onlinecite{MacDonaldGirvin88} that the unitary transformation
can be written as
\bea
e^{-iS} &=& \sum_{k\geq 0} \sum_{[m]} \left(\frac{t}{U}\right)^k
\alpha^{(k)}[m] T^{(k)}[m],
\label{eq:unitaryExpansion}\\
T^{(k)}[m] &=& T_{m_1} T_{m_2} \dots T_{m_k}\ ,\quad m_j\in\{-1,0,1\},
\eea
where $\alpha^{(k)}[m]$ are suitably chosen coefficients.

%%%%%%%%%%%%%%%%%%%%%%%%%%%%%%%%%%%%%%%%%%
\subsection{Heisenberg sector states in the $t/U$ expansion}
\begin{figure}[ht]
\includegraphics[width=0.45\textwidth]{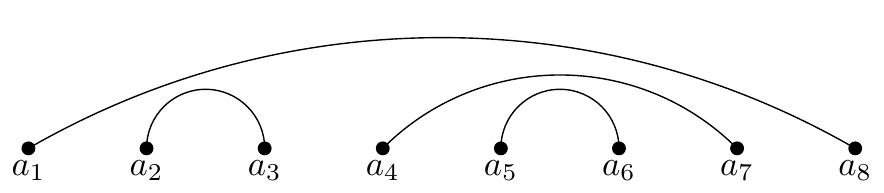}
\caption{The pairing
  $\boldsymbol{k}=\{a_1,a_8,a_2,a_3,a_4,a_7,a_5,a_6\}$ does not
  involve crossed lines and fulfils the constraint.}
\label{fig:nocrossing}
\end{figure}

%%%%%%%%%%%%%%%%%%%%%%%%%%%%%%%%%%%%%%%%%%
%\emph{The ``Heisenberg sector''.}
We now need to identify the Heisenberg sector states in the framework
of the $t/U$-expansion. We propose that they are characterized by
their property that, in the framework of the $t/U$-expansion, they are
connected by our unitary transformation to states without any double
occupancies
\be
|\psi_H\rangle\Big|_{t/U}=
|\psi\rangle=e^{-iS}\ \sum_{\alpha_j=\up,\down} 
f_{\alpha_1\dots\alpha_L}\Big[\prod_{j=1}^LX^{\alpha_j0}\Big]|0\rangle.
\label{psiH}
\ee
Our identification is based on the fact that in the Bethe ansatz
solution exact eigenstates are labelled by sets of (half-odd) integer
quantum numbers, which makes it possible to follow a particular state
when changing the interaction strength $U$. When sending $U/t$ to infinity
for a large but fixed system size one finds that in this limit the
lowest energy eigenstates belong to the Heisenberg sector, which in
turn leads to the identification \fr{psiH}.
In the following we will use the eigenstate $|\psi_H\rangle$ and its
$t/U$-expansion $|\psi\rangle$ interchangeably and leave a discussion
of contributions not captured by the $t/U$-expansion for our
concluding remarks. The
ground state of the half-filled Hubbard model is a particular example
of a Heisenberg state. Heisenberg sector states have the important
property that they are singlets under the $\eta$-pairing SU(2)
algebra. This follows from the easily established fact that 
\be
[\eta,T_m]=0=[\eta^\dagger,T_m]=[\eta^z,T_m]\ ,\quad m=0,\pm 1.
\label{etaT}
\ee
The commutation relations \fr{etaT} imply that $T^{(k)}[m]$ commute
with the $\eta$-pairing operators, and this in turn implies by virtue of
\fr{eq:unitaryExpansion} that
\be
[\eta,e^{-iS}]=0=[\eta^\dagger,e^{-iS}]=[\eta^z,e^{-iS}].
\ee
This, together with the fact that all states with only singly occupied
sites are annihilated by both $\eta$ and $\eta^\dagger$, establishes
that all Heisenberg states are $\eta$-pairing singlets
\be
\eta|\psi\rangle=\eta^\dagger|\psi\rangle=\eta^z|\psi\rangle=0.
\ee
Using the results of Refs~\onlinecite{temperleyLieb,saitoproof} we may
construct a basis for the space of $\eta$-pairing singlets. Let us
select a set $A=\{a_1,\dots,a_{2q}\}$ of $2q$ lattice sites and denote
the  complementary set by $\bar{A}$. We take all $\bar{A}$ sites to be
singly occupied by spin-$\sigma_j$ fermions, while we form singlet
$\eta$-pairing dimers $|S(a,b)\rangle=|2\rangle_{a}
  |0\rangle_{b} + (-1)^{a+b - 1} |0\rangle_{a} |2\rangle_{b}$
on the $A$ sites. This gives an overcomplete set
of states 
\bea
|\boldsymbol{k};\boldsymbol{\sigma}\rangle=\prod_{i=1}^{q}
|S(k_{2i-1},k_{2i})\rangle\prod_{j\in\bar{A}}|\sigma_j\rangle_j,
\label{basis}
\eea
where $\boldsymbol{k}=k_1,\dots,k_{2q}$ is a permutation of
$\{a_1,\dots,a_{2q}\}$. It is
  easily verified that the states 
\fr{basis} are $\eta$-pairing singlets. A linearly independent set of
states \fr{basis} is formed by imposing a ``non-crossing'' constraint
on the permitted values of $\boldsymbol{k}$ as follows. We first
impose an ordering of sites $\{a_1,\dots,a_{2q}\}$,
e.g. $a_1<a_2<\dots<a_{2q}$. In one dimension this is simply the
natural order of the sites. We then connect the $q$ pairs of sites
$\{(k_{2i-1},k_{2i})\}$ by lines, \emph{cf.} Fig.~\ref{fig:nocrossing}. If
no lines cross, the state corresponding to the pairing
$\boldsymbol{k}$ is permitted. A basis $\mathfrak{B}$ of all
$\eta$-pairing singlet states in the half-filled Hubbard model is 
obtained by taking into account all ``sectors'' $0\leq 2q\leq L$ and
all distinct sets $\{a_1,\dots,a_{2q}\}$ of lattice sites in a given
sector. 
All eigenstates in the Heisenberg sector can then be expressed in the form
\be
|\psi\rangle=\sum_{|\boldsymbol{k};\boldsymbol{\sigma}\rangle\in\mathfrak{B}}
A_{\boldsymbol{k};\boldsymbol{\sigma}}\
|\boldsymbol{k};\boldsymbol{\sigma}\rangle\ .
\label{Heisstate}
\ee
%%%%%%%%%%%%%%%%%%%%%%%%%%%%%%%%%%%%%
\subsection{One spatial dimension} 
%%%%%%%%%%%%%%%%%%%%%%%%%%%%%%%%%%%%%
The $t/U$-expansion allows us to
determine the dependence of the amplitudes
$A_{\boldsymbol{k};\boldsymbol{\sigma}}$ on $U$. It is useful to
define the ``total bond length'' by 
\bea
\mathcal{D}\Big(
\sum_{|\boldsymbol{k};\boldsymbol{\sigma}\rangle\in\mathfrak{B}}
A_{\boldsymbol{k};\boldsymbol{\sigma}}\
|\boldsymbol{k};\boldsymbol{\sigma}\rangle\Big)
&=&\max_{\ontop{|\boldsymbol{k};\boldsymbol{\sigma}\rangle\in\mathfrak{B}}
{A_{\boldsymbol{k};\boldsymbol{\sigma}}\neq 0}}
 \mathcal{D}\left(|\boldsymbol{k};\boldsymbol{\sigma}\rangle\right),
\label{eq:bondDistDef}
\eea
where we take $\mathcal{D}\big(|\boldsymbol{k};\boldsymbol{\sigma}\rangle\big)
=\sum_{i=1}^q (\lVert k_{{2i-1}} - k_{{2i}} \rVert +1)$.
It is a straightforward matter to show (see the Supplementary
Material) that
$\mathcal{D}\big(T_n |\psi\rangle \big) \leq 
\mathcal{D}\big(|\psi\rangle \big) +n+1$ where $n=0,\pm 1$ and
$|\psi\rangle$ is any Heisenberg sector state \fr{Heisstate}. This in
turn implies that $\mathcal{D}\left[ T^{(k)}[m] |\psi_0\rangle \right] 
\leq k+q$, where $q=\sum_{i=1}^k m_i$ and $|\psi_0\rangle$ is any
state with only singly occupied sites. Applying this to the expressions
\fr{psiH}, \fr{eq:unitaryExpansion} for Heisenberg states, we conclude
that the expansion coefficients in \fr{Heisstate} fulfil
\be
A_{\boldsymbol{k};\boldsymbol{\sigma}}={\cal O}\Big(
\big(t/U\big)^{\sum_{i=1}^q \lVert k_{{2i}} - k_{{2i-1}}\rVert} \Big).
\label{eq:UDep}
\ee
These results cannot be straightforwardly generalized to $D>1$ because
our definition of a total bond length hinges on
$|\boldsymbol{k};\boldsymbol{\sigma}\rangle$ forming a basis of
states, which imposes constraints on the allowed values of
$\boldsymbol{k}$. In a typical Heisenberg sector state we have a
finite density of doubly occupied sites and the coefficients
\fr{eq:UDep} are of an extremely high order in $t/U$. In order to
proceed we will assume that the $t/U$-expansion for the wave-function
\fr{Heisstate} has a finite radius of convergence. We know this to be
the case for certain quantities such as the ground state energy
\cite{book}.

%%%%%%%%%%%%%%%%%%%%%%%%%%%%%%%%%%%%%%%%%%%
\subsection{Quantum disentangled diagnostic}
%%%%%%%%%%%%%%%%%%%%%%%%%%%%%%%%%%%%%%%%%%%
%%%%%%%%%%%%%%%%%%%%%%%%%%%%%%%%%%%%%%%%%%%
%\emph{Quantum disentangled diagnostic.}
%%%%%%%%%%%%%%%%%%%%%%%%%%%%%%%%%%%%%%%%%%%
According to Refs~\onlinecite{qdl,hubb1} a QDL can be diagnosed by
preparing the system in a finite energy-density eigenstate with
volume-law bipartite EE, and then to carry out a
projective measurement of the $z$-component of spin on each site of
the lattice. If the resulting state is characterized by an area-law
EE, the original state realizes a QDL.

\begin{figure}[ht]
\includegraphics[width=0.4\textwidth]{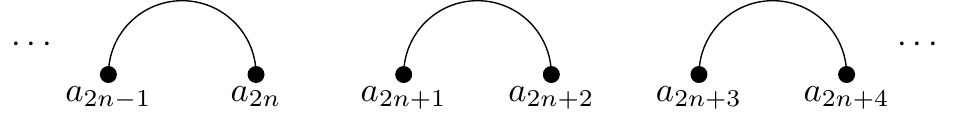}
\caption{Structure of the singlet pairings in the leading 
non-vanishing term of the projected state in the $t/U$-expansion.}
\label{fig:leading}
\end{figure}

We now address this proposal in the framework of the $t/U$-expansion.
As our initial state we choose a finite energy density eigenstate
\fr{psiH} in the Heisenberg sector. These generically have 
volume-law entanglement entropies as can be seen from the fact that
the corresponding macro-states have finite thermodynamic entropy
densities. As the 1D Hubbard model is integrable there also exist
finite energy density eigenstates with area-law EE, but these are the
exception rather than the rule. Let us assume that the outcome of our
projective spin measurement is that we obtain spin zero at all sites
in the set $A=\{a_1,a_2,\dots a_{2q}\}$ and spin $\sigma_j=\pm 1/2$
everywhere else. Then the state of the system after the projective
measurement can be written as  
\bea
|\psi_{\rm proj}\rangle &=& 
\frac{1}{\sqrt{\mathcal{N}}}
\prod_{j_1=1}^{a_1-1} X_{j_1}^{\sigma_{j_1} \sigma_{j_1}}
\left( X_{a_1}^{00} + X_{a_1}^{22} \right)\nn
&\times&\prod_{j_2=a_1+1}^{a_2-1} 
X_{j_2}^{\sigma_{j_2} \sigma_{j_2}}
\left( X_{a_2}^{00} + X_{a_2}^{22} \right) \cdots
|\psi\rangle% = P|\psi\rangle
, \label{eq:projDef}
\eea
where $\mathcal{N}$ is a normalisation factor. Using our results
\fr{Heisstate} for the structure of Heisenberg states in
the $t/U$ expansion, we can rewrite this in the form
\be
|\psi_{\rm proj}\rangle = \sum_{Q\in S_{2q}}' W(Q) \prod_{i=1}^{q}
|S(k_{Q_{2i-1}},k_{Q_{2i}})\rangle
|\boldsymbol\sigma\rangle. \label{eq:bellconjecture}
\ee
Here $|\boldsymbol\sigma \rangle = \prod_{j \notin A}
|\sigma_j\rangle$ is a product state fixed by the projective
measurement, and the sum $\sum'_{Q\in S_{2q}}$ is restricted to the
permutations $Q$ such that $a_{Q_1},a_{Q_2}, \dots a_{Q_{2q}}$
corresponds to singlet states that satisfy the non-crossing
condition. As a result of \fr{eq:UDep} the amplitudes $W(Q)$ fulfil
\be
W(Q) \sim \mathcal{O}\left( \big(t/U\big)^{\sum_{i=1}^q
\lVert k_{Q_{2i}} - k_{Q_{2i-1}}\rVert} \right).
\ee
In the leading order in the $t/U$-expansion that is allowed to be
non-zero by the above considerations there is only a single term in
\fr{eq:bellconjecture} and we are dealing with a spatially ordered
product state of singlet dimers, 
\emph{cf.} Fig.~\ref{fig:leading}.
In the following we will assume for simplicity that the numerical
coefficient of this term is indeed different from zero, which will be
generically the case. Depending on the choice of the initial
Heisenberg state and the outcome of the projective measurement it is
possible that this coefficient vanishes. Such cases can be
accommodated by relatively straightforward modifications of the
following discussion.

%%%%%%%%%%%%%%%%%%%%%%%%%%%%%%%%%%%%%%
\subsection{Entanglement entropy.}
\begin{figure}[ht]
\includegraphics[width=0.45\textwidth]{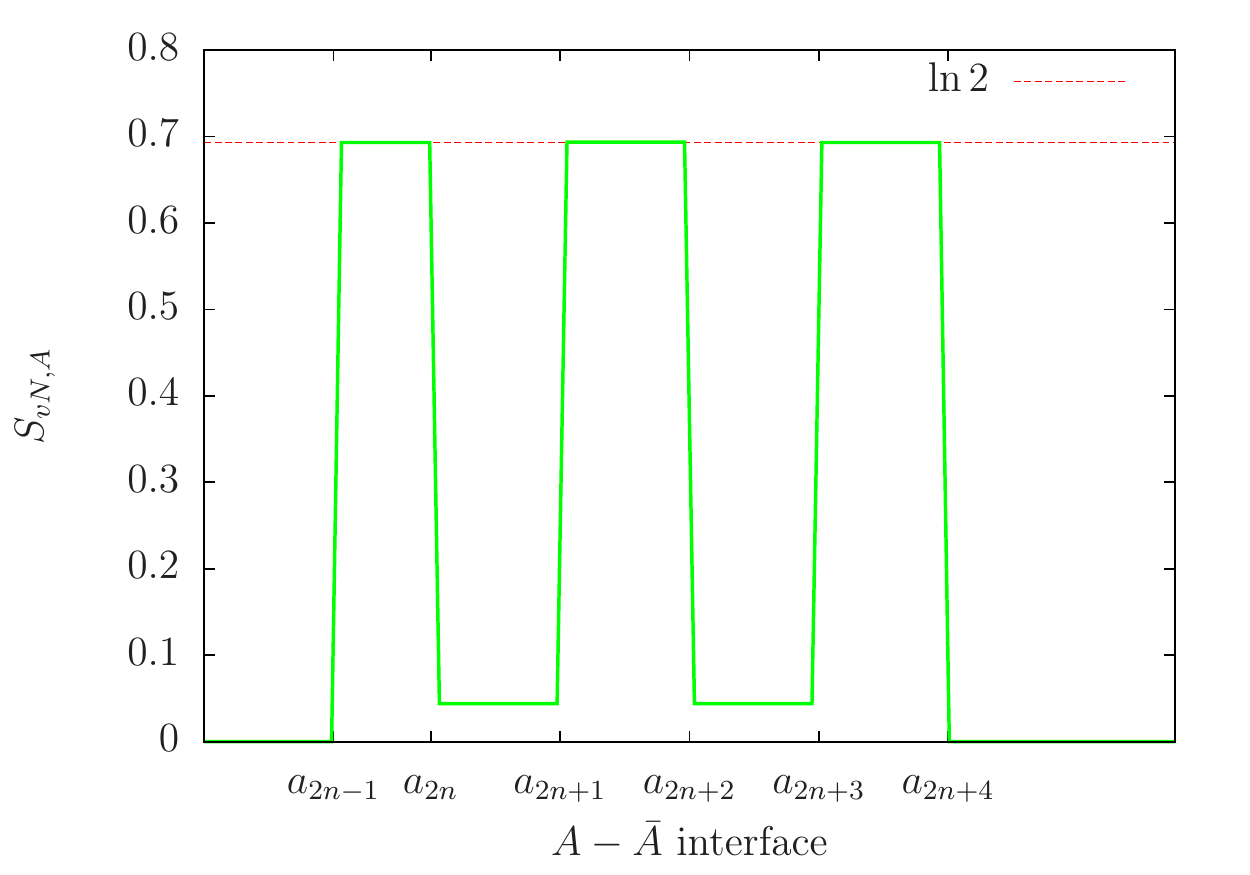}
\caption{Form of the bipartite entanglement entropy after projective
measurement of a general Heisenberg state. Deviations from $S(x)=\ln 2, 0$
arise from higher order terms, as seen in (\ref{eq:higherOrderTerms}),
 (\ref{eq:higherOrderCorrections}).}
\label{fig:sketch2}
\end{figure}

%%%%%%%%%%%%%%%%%%%%%%%%%%%%%%%%%%%%%%
%%%%%%%%%%%%%%%%%%%%%%%%%%%%%%%%%%%%%%
%\emph{Entanglement entropy.}
%%%%%%%%%%%%%%%%%%%%%%%%%%%%%%%%%%%%%%
At any finite order in the $t/U$-expansion the projected states
\fr{eq:bellconjecture} are only weakly entangled. Let us consider the
generic case, in which the leading non-zero term in the projected
state has the form shown in Fig.~\ref{fig:leading}. In this case
the leading term in von Neumann entropy of a sub-system $A$ is
simply given by $S_{{\rm vN},A}=\ln(2)$ if we cut one of the singlet
dimers when we bipartition the system, and zero otherwise. At leading
order it is not possible for the bipartition to cut more than one
dimer. In order to describe the structure of the leading corrections
to this result we take the subsystem $A$ to be the interval
$[1,\ell]$, where $a_{2n+1}<\ell<a_{2n+2}$. Here $\{a_j\}$ are the
sites in the projected state that are either doubly occupied or empty.
We then cast the projected state in the form
\begin{widetext}
\beA
|\psi_{\rm proj}\rangle &= 
|\psi_L\rangle\left(
\Bigg\vert                                                                      
\begin{tikzpicture}[scale=0.9,baseline=-.2ex]                                   
	\filldraw (-.5,0) node[below] {$a_{2n-1}$} circle (2pt); 
	\filldraw (0.5,0) node[below] {$a_{2n}$} circle (2pt); 
	\filldraw (1.5,0) node[below] {$a_{2n+1}$} circle (2pt); 
	\filldraw (2.5,0) node[below] {$a_{2n+2}$} circle (2pt); 
	\filldraw (3.5,0) node[below] {$a_{2n+3}$} circle (2pt); 
	\filldraw (4.5,0) node[below] {$a_{2n+4}$} circle (2pt); 
	\draw (-.5,0) arc (180:0:.5);
	\draw (1.5,0) arc (180:0:.5);
	\draw (3.5,0) arc (180:0:.5);
	\draw[dashed] (2,-.5) -- (2,1.0);
\end{tikzpicture}                                                               
\Bigg\rangle 
+
\varepsilon_1
\Bigg\vert                                                                      
\begin{tikzpicture}[scale=0.9,baseline=-.2ex]                                   
	\filldraw (-.5,0) node[below] {$a_{2n-1}$} circle (2pt); 
	\filldraw (0.5,0) node[below] {$a_{2n}$} circle (2pt); 
	\filldraw (1.5,0) node[below] {$a_{2n+1}$} circle (2pt); 
	\filldraw (2.5,0) node[below] {$a_{2n+2}$} circle (2pt); 
	\filldraw (3.5,0) node[below] {$a_{2n+3}$} circle (2pt); 
	\filldraw (4.5,0) node[below] {$a_{2n+4}$} circle (2pt); 
	\draw (-.5,0) arc (180:0:.5);
%	\draw (1.5,0) arc (180:0:1.5);
	\draw (1.5,0) arc (135:45:2.121);
	\draw (2.5,0) arc (180:0:.5);
	\draw[dashed] (2,-.5) -- (2,1.0);
\end{tikzpicture}                                                               
\Bigg\rangle 
\right.\\
&
\left.
+
\varepsilon_2
\Bigg\vert                                                                      
\begin{tikzpicture}[scale=0.9,baseline=-.2ex]                                   
	\filldraw (-.5,0) node[below] {$a_{2n-1}$} circle (2pt); 
	\filldraw (0.5,0) node[below] {$a_{2n}$} circle (2pt); 
	\filldraw (1.5,0) node[below] {$a_{2n+1}$} circle (2pt); 
	\filldraw (2.5,0) node[below] {$a_{2n+2}$} circle (2pt); 
	\filldraw (3.5,0) node[below] {$a_{2n+3}$} circle (2pt); 
	\filldraw (4.5,0) node[below] {$a_{2n+4}$} circle (2pt); 
%	\draw (-.5,0) arc (180:0:1.5);
	\draw (-.5,0) arc (135:45:2.121);
	\draw (0.5,0) arc (180:0:0.5);
	\draw (3.5,0) arc (180:0:.5);
	\draw[dashed] (2,-.5) -- (2,1.0);
\end{tikzpicture}                                                               
\Bigg\rangle \right)|\psi_R\rangle+\dots
,
\label{eq:higherOrderTerms}
\eeA
\end{widetext}
where $|\psi_L\rangle$ ($|\psi_R\rangle$) is the part of the
leading term of the projected state that involves the sites
$j<a_{2n-1}$ ($j>a_{2n+4}$). The coefficients $\epsilon_{1,2}$
are of order $(t/U)^{2(a_{2n+3}-a_{2n+2})}$ and
$(t/U)^{2(a_{2n+1}-a_{2n})}$ respectively. The von Neumann entropy for
our subsystem is then
\be
S_{{\rm vN},A}= \ln 2 + \frac{9}{16}\varepsilon_1^2\varepsilon_2^2
\big( 
1 + 2 \ln\Big( \frac{4}{3} \Big) - 2 \ln(\varepsilon_1\varepsilon_2)
\big)
+\dots
\label{eq:higherOrderCorrections}
\ee
We note that the correction is positive. The physical picture that
emerges is very simple: for small $t/U$, the
projected state is very close to being a product of Bell pairs and
the bipartite EE takes  the form shown in
Fig.~\ref{fig:sketch2}.
The average double occupancy in a Heisenberg state is ${\cal
  O}(t^2/U^2)$, which implies that the average distance between doubly
occupied/unoccupied sites is very large $\sim U^2/t^2$. Hence, if the
outcome of our spin measurement is close to its average, the sites
$a_1,\dots, a_{2q}$ are well separated and the leading term
\fr{fig:leading} is expected to provide an excellent approximation as
long as the $t/U$-expansion converges. We note that the results
obtained by the $t/U$-expansion are fully compatible with those
obtained by integrability methods in section
\ref{sec:integrability}. This, and the fact that the $t/U$-expansion
is known to converge for simple quantities like the ground state
energy\cite{book,takahashigsE}, provides support for its
applicability.

%%%%%%%%%%%%%%%%%%%%%%%%%%%%%%%%%%%%%%%%%%
\subsection{Higher dimensions $D>1$}
%%%%%%%%%%%%%%%%%%%%%%%%%%%%%%%%%%%%%%%%%%
%\emph{Higher dimensions $D>1$.} 
A key element of our analysis in the one
dimensional case is the notion of a bond length, which fulfils
${\cal D}(|\psi_1\rangle)\leq{\cal D}(|\psi_1\rangle+|\psi_2\rangle)$
for arbitrary $\eta$-pairing singlet states $|\psi_{1,2}\rangle$. Our
definition of a bond length utilizes the availability of a convenient
basis of such states. It is this aspect which makes the
$D=1$ special. In $D>1$ we proceed by again first fixing a set
$A=\{a_i\}$ of $2q$ unoccupied or doubly occupied sites. We denote by
$E_{ij}$ a bond that joins sites $a_i$ and $a_j$. We then define
a dimer configuration $\mathfrak{C}$ as a set of $q$ bonds $C_{ij}$
connecting sites $a_i$ and $a_j$ such that all sites belong
to precisely one bond. Each dimer configuration gives rise to
$\eta$-pairing singlet states of the form 
\be
|\mathfrak{C};{\boldsymbol\sigma}\rangle = 
\prod_{ C_{ij} \in \mathfrak{C} } |S(a_{i},a_{j})\rangle
\prod_{j\notin A}|\sigma_j\rangle_j\ .
\label{dimercoverings}
\ee
The states \fr{dimercoverings} form an overcomplete set.
In order to select a set of linearly independent states we use the
function $\overline{\mathcal{D}}$ defined by
%\be
$\overline{\mathcal{D}}\,(|\mathfrak{C};{\boldsymbol\sigma}\rangle ) = 
\sum_{C_{ij}\in\mathfrak{C}} \big(\lVert a_{i} - a_{j} \rVert + 1\big)$,
%\ee
where $\lVert a_{i} - a_{j} \rVert$ denotes the Manhattan distance
between sites $a_i$ and $a_j$. Among \fr{dimercoverings} we
select the $\frac{(2q)!}{(q+1)!(q!)}$ states that have the lowest
values under the map $\overline{\mathcal{D}}$: it is this criterion that 
allows the inequalities in 1D to be inherited in the higher-dimensionsal 
case.
Repeating this construction for all values of $2q\leq L$ and all distinct 
sets $A$ gives a basis $\mathcal{B}$ of $\eta$-pairing singlet states.
We now define our total bond distance $\mathcal{D}$ in the
same way as in (\ref{eq:bondDistDef}) with the covering $\mathfrak{C}$
taking the place of the vector $\boldsymbol{k}$. With these
definitions in place it is straightforward to show (see the
Supplementary Material) that $\mathcal{D}\big(T_n |\psi\rangle \big)
\leq  \mathcal{D}\big(|\psi\rangle \big) +n+1$ where $n=0,\pm 1$ and
$|\psi\rangle$ is any Heisenberg sector state 
\be
|\psi\rangle=\sum_{|\mathfrak{C};\boldsymbol{\sigma}\rangle\in\mathfrak{B}}
A_{\mathfrak{C};\boldsymbol{\sigma}}\
|\mathfrak{C};\boldsymbol{\sigma}\rangle\ .
\label{Heisstate2D}
\ee
This in turn implies that $\mathcal{D}\left[ T^{(k)}[m] |\psi_0\rangle
  \right]  \leq k+q$, where $q=\sum_{i=1}^k m_i$ and $|\psi_0\rangle$
is any state with only singly occupied sites. It follows that the
$d$-dimensional generalisation of (\ref{eq:UDep}) is
\be
A_{\mathfrak{C};\boldsymbol{\sigma}}={\cal O}\Big(
\big(t/U\big)^{\sum_{C_{ij}\in\mathfrak{C}} \lVert a_i - a_j\rVert} \Big).
\ee
The main difference to $D=1$ is that now there can be several basis
states contributing to the leading order term in $t/U$. To see this
we may consider a square lattice and focus on the situation where the
sites in $A$ completely fill an $m\times n$ rectangle (with at least
one of $m$, $n$ even). The leading-order term is that generated by
$(T_1)^{mn/2}$ produces a superposition of all states in $\mathfrak{B}$
that correspond to nearest-neighbour dimer coverings of this
rectangle. Nevertheless, it is apparent that to leading order in the
$t/U$-expansion the structure of the state is such that the
entanglement entropy follows an area-law. 

%%%%%%%%%%%%%%%%%%%%%%
\section{Conclusions}
\label{sec:conc}
%%%%%%%%%%%%%%%%%%%%%%
We have shown that there are particular ``Heisenberg sector''
eigenstates in the one dimensional half-filled Hubbard model that
realize the quantum disentangled liquid state of matter in the strong
sense proposed in Ref.~\onlinecite{qdl}. These states are obtained by
freezing the charge degrees of freedom in their ground state
configuration in the framework of the Bethe ansatz solution of the
Hubbard model. Using methods of integrability we have demonstrated
that the charge degrees of freedom (``light particles'') do not
contribute to the volume term of the bipartite entanglement
entropy. Employing strong-coupling expansion techniques we have shown
(under the assumption that the expansion converges) that a measurement
of the spin degrees of freedom at all sites (``heavy particles'')
leaves the system in a state that is area-law entangled, which is the
defining characteristic of a ``strong QDL''\cite{qdl}. 

In contrast to Heisenberg sector states, the entanglement entropy for
maximal entropy (thermal) states at a given energy density at large
values of $U/t$ does have a contribution that involves the charge
degrees of freedom, but it is of the form \fr{weak}, i.e.
\bea
S_{\rm vN,A}&=&\big(s_{\rm spin}+s_{\rm
  charge}\big)|A|+o\big(|A|\big)\ ,\nn
s_{\rm spin}&=&{\cal O}(1)\ ,\quad
s_{\rm charge}={\cal O}\Big(\frac{u}{T}e^{-u/T}\Big)\ ,\quad
\eea
We expect a similar volume law to occur in the entanglement entropy
after a measurement of all spins. This suggests that there is a \emph{weak}
variant of a QDL, which is characterized by a volume term in the EE
after measurement of the heavy degrees of freedom that is
exponentially small in the ratio of masses (i.e. the ratio $U/t$ in
the Hubbard model). In this limit the volume term is only observable 
in enormously (exponentially) large systems and would be practically
impossible to detect in numerical simulations. We expect this
weak scenario to be realized quite generally in strong coupling limits
irrespective of whether the model one is dealing with is integrable or
not. In particular, this scenario is compatible with our
strong-coupling analysis of the QDL diagnostic in $D=2$.

An interesting question is how to access Heisenberg sector states in
the one dimensional case. In principle this can be achieved by means
of a quantum quench \cite{EFreview} from a suitably chosen initial
state. At late times the system locally relaxes to a steady state that
is described by an appropriate generalized Gibbs ensemble. The
parameters of this ensemble are fixed by the initial conditions and
can in principle be tuned in such a way that one ends up with a
Heisenberg sector state. How to do this in practice is an interesting
albeit rather non-trivial question. Fixing the energy density to be
small compared to the Mott gap is sufficient to obtain a
non-equilibrium steady state that realizes the weak form of a QDL
characterized by \fr{weak}. In order to remove the $s_{\rm charge}$
contribution, the expectation values of higher conservation laws in
the initial state have to be chosen appropriately and it would be
interesting to investigate this issue further. 

\acknowledgments
We are grateful to P. Fendley, J. Garrison, T. Grover, L. Motrunich and M.
Zaletel for helpful discussions. This work was supported by the EPSRC under
grant EP/N01930X (FHLE), by the National Science Foundation,
under Grant No. DMR-14-04230 (MPAF) and by the Caltech Institute of
Quantum Information and Matter, an NSF Physics Frontiers Center with
support of the Gordon and Betty Moore Foundation (MPAF). 

\appendix
%%%%%%%%%%%%%%%%%%%%%%%%%%%%%%%%%%%%%%%
\section{Counting Heisenberg sector states}
\label{app:counting}
%%%%%%%%%%%%%%%%%%%%%%%%%%%%%%%%%%%%%%%
In order to get a measure of how many Bethe ansatz Heisenberg sector
states there are we proceed as follows. We start with the
thermodynamic limit of the Bethe ansatz equations \fr{densities} for
Heisenberg sector states
\begin{eqnarray}
\rho^p(k)&=&\frac{1}{2\pi}+\cos
k\sum_{n=1}^\infty\int_{-\infty}^\infty d\Lambda\ \fa{n}{\l-\sin k}
\sigma_n^p(\Lambda)\ ,\nn
\sigma_n^h(\Lambda)&=&-\sum_{m=1}^\infty
\int_{-\infty}^\infty d\Lambda'
A_{nm}(\Lambda-\Lambda')\ \sigma_m^p(\Lambda')\nn
&+&\int_{-\pi}^\pi dk\ \fa{n}{\sin k-\Lambda}\ \rho^p(k)\ .
\label{densitiesHS}
\end{eqnarray}
We then construct the maximum entropy state at a given energy density
by extremizing the free energy
\be
f=e-{\cal T}s-B\left[\int_{-\pi}^\pi dk\ \rho^p(k)
-\sum_{n=1}^\infty2n\int_{-\infty}^\infty d\Lambda \sigma^p_n(\Lambda)
\right]\ ,
\label{eff}
\ee
where $e$ and $s$ are the energy and entropy densities of Heisenberg
sector states and are given by \fr{entropy} and \fr{SH}
respectively. $B$ plays the role of a magnetic field and will
eventually be set to zero. The ``temperature'' ${\cal T}$ plays the
role of a Lagrange parameter that allows us to fix the energy density.
Extremizing \fr{eff} with respect to the particle and hole densities
under the constraints \fr{densitiesHS} fixes the ratios
$\eta_n=\frac{\sigma^h_n}{\sigma^p_n}$ for the maximal entropy state
at a given value of ${\cal T}$ via the TBA-like equations
\bea
&&\ln\big(1+\eta_n(\Lambda)\big)=
\frac{g_1(\Lambda)}{{\cal T}}\nn
&&+\sum_{m=1}^\infty\int_{-\infty}^\infty d\Lambda'\
A_{nm}(\Lambda-\Lambda')\ln\left[1+\frac{1}{\eta_n(\Lambda')}\right],\ \
\label{TBAHS}
\eea
where $g_1(\Lambda)=-4{\rm     Re}\sqrt{1-(\Lambda-inu)^2}+4nu+2nB$.
The entropy density of this state is
\be
s=\sum_{n=1}^\infty\int_{-\infty}^\infty d\Lambda\left[
\frac{g_1(\Lambda)}{{\cal T}}\sigma_n^p(\Lambda)+
g_2(\Lambda)\ln\big(1+\eta_n^{-1}(\Lambda)\big)
\right],
\ee
where $g_2(\Lambda)=\frac{1}{\pi}{\rm Re}\frac{1}{\sqrt{1-(\Lambda+inu)^2}}$.
In order to obtain a thermodynamic estimate of the number of
Heisenberg sector states we now consider the limit $B,{\cal
  T}\to\infty$ at fixed $B/{\cal T}$. In this limit the solution of
the system of equations \fr{TBAHS} is\cite{book}
\be
\eta_n=\left[\frac{\sinh\Big(\frac{(n+1)B}{{\cal
        T}}\Big)}{\sinh\big(\frac{B}{{\cal T}}\big)}\right]^2-1.
\ee
Finally we take $B/{\cal T}$ to zero
\be
\eta_n=(n+1)^2-1.
\ee
Substituting this back into the expression \fr{TBAHS} for the entropy
we have
\be
\lim_{{\cal T}\to\infty}s=\sum_{n=1}^\infty\ln\left[\frac{(n+1)^2}{(n+1)^2-1}\right]=\ln(2).
\ee
This tells us that the most likely macro state in the Heisenberg
sector has an entropy density of $\ln(2)$. All other macro states in
this sector are exponentially less likely, and hence we conclude that
the total number of Heisenberg sector micro states fulfils \fr{HScount}.

%%%%%%%%%%%%%%%%%%%%%%%%%%%%%%%%%%%%%%%
\section{Thermodynamic Bethe Ansatz equations}
\label{app:TBAeqns}
%%%%%%%%%%%%%%%%%%%%%%%%%%%%%%%%%%%%%%%
The Thermodynamic Bethe Ansatz equations for the one dimensional
Hubbard model in zero magnetic field are\cite{takahashiTBA1,book}
\begin{widetext}
\begin{eqnarray}
\ln \zeta(k)&=&\frac{-2\cos k -\mu-2u}{T}
+\sum_{n=1}^\infty 
\int_{-\infty}^\infty 
d\Lambda\ \fa{n}{\sin k-\Lambda}
\ln\left(1+\frac{1}{\eta^\prime_n(\Lambda)}\right)\nn
&-&\sum_{n=1}^\infty \int_{-\infty}^\infty
d\Lambda\ \fa{n}{\sin k-\Lambda}
\ln\left(1+\frac{1}{\eta_n(\Lambda)}\right),
\label{zeta}\\
\ln\left(1+\eta_n(\Lambda)\right) &=& -\int_{-\pi}^\pi dk 
\ \cos(k)\ \fa{n}{\sin k-\Lambda}
\ln\left(1+\frac{1}{\zeta(k)}\right)
+\sum_{m=1}^\infty
A_{nm}*\ln\left(1+\frac{1}{\eta_m}\right)\bigg|_\Lambda\ ,
\label{lambda}\\
%\end{eqnarray}
%\begin{eqnarray}
\ln\left(1+\eta^\prime_n(\Lambda)\right)&=&
\frac{4{\rm Re}\sqrt{1-(\Lambda -inu)^2}-2n\mu-4nu}{T}
-\int_{-\pi}^\pi dk \ \cos(k)\ \fa{n}{\sin k-\Lambda}
\ln\left(1+\frac{1}{\zeta(k)}\right)\nn
&+&\sum_{m=1}^\infty A_{nm}*\ln\left(1+\frac{1}{\eta^\prime_m}\right)
\bigg|_\Lambda .
\label{kl}
\end{eqnarray}
\index{TBA equations}
\end{widetext}

%%%%%%%%%%%%%%%%%%%%%%%%%%%%%%%%%%%
\section{Some inequalities for the total bond length of Heisenberg states}
\label{app:bondInequalities}
%%%%%%%%%%%%%%%%%%%%%%%%%%%%%%%%%%%
%%%%%%%%%%%%%%%%%%%%%%%%%%%%%%%%%%%
%\section*{Supplementary Material}
%\label{app:bondInequalities}
%%%%%%%%%%%%%%%%%%%%%%%%%%%%%%%%%%%
Here we show that
\be
\mathcal{D}\left( T_n|\psi\rangle \right) \leq \mathcal{D}\left( |\psi\rangle
\right) + n +1,\qquad n=0,\pm1.
\label{ineq}
\ee
We consider the one and higher dimensional cases separately.
%%%%%%%%%%%%%%%%%%%%%
\subsection{One Dimension}
%%%%%%%%%%%%%%%%%%%%%
Any Heisenberg state can be written in the form \fr{Heisstate}
\be
|\psi\rangle=\sum_{|\boldsymbol{k};\boldsymbol{\sigma}\rangle\in\mathfrak{B}}
A_{\boldsymbol{k};\boldsymbol{\sigma}}\
|\boldsymbol{k};\boldsymbol{\sigma}\rangle\ .
\ee
To establish \fr{ineq} it therefore suffices to show that
\be
\mathcal{D}\left( T_n |{\bf k};\boldsymbol{\sigma}\rangle \right) \leq
\mathcal{D}\left( |{\bf k};\boldsymbol{\sigma}\rangle\right) + n + 1\ .
\ee
We next consider the three cases $n=0,\pm 1$ in turn.
%%%%%%%%%%%%%%%%%%%%%%%%%%%%%%%%%
\subsubsection{Inequality for $T_1$:}
%%%%%%%%%%%%%%%%%%%%%%%%%%%%%%%%%
Due to the definition of the bond distance, it is clear that
\beA
\mathcal{D}
\left( \sum_j T_{m,j}|{\bf k};{\boldsymbol\sigma}\rangle \right)
&\leq
\sup_j 
\mathcal{D}
\left(  T_{m,j} |{\bf k};{\boldsymbol\sigma}\rangle \right),
\eeA
and therefore for $m=\pm1,0$, we simply need to find the case which yields the
largest value in order to prove the inequality.
$T_{1,j}$ creates an $\eta$-singlet on two adjacent sites and $\mathcal{D}$
counts the total number of sites spanned by the singlet bonds, applying
$T_{1,j}$ increases this by at most that in the case of a system with no doubly
occupied sites. We can view the result graphically
\beA
\mathcal{D}\left( T_{1,j} 
\Bigg\vert                                                                      
\begin{tikzpicture}[scale=0.5, every node/.style={scale=0.55},baseline=-.2ex]                                   
	\filldraw (1.5,0) node[below] {$a_{2n-2}$} circle (2pt); 
	\filldraw (2.5,0) node[below] {$a_{2n-1}$} circle (2pt); 
	\draw (1.5,0) arc (180:0:.5);
\end{tikzpicture}                                                               
\Bigg\rangle 
\right)
&
\leq
\mathcal{D}\left( 
\Bigg\vert                                                                      
\begin{tikzpicture}[scale=0.5, every node/.style={scale=0.55},baseline=-.2ex]                                   
	\filldraw (1.5,0) node[below] {$a_{2n-2}$} circle (2pt); 
	\filldraw (2.5,0) node[below] {$a_{2n-1}$} circle (2pt); 
	\filldraw (3.5,0) node[below] {$j$} circle (2pt); 
	\filldraw (4.5,0) node[below] {$j+1$} circle (2pt); 
	\draw (1.5,0) arc (180:0:.5);
	\draw (3.5,0) arc (180:0:.5);
\end{tikzpicture}                                                               
\Bigg\rangle 
\right),\\
&
\leq
\mathcal{D}\left(
\Bigg\vert                                                                      
\begin{tikzpicture}[scale=0.5, every node/.style={scale=0.55},baseline=-.2ex]                                   
	\filldraw (1.5,0) node[below] {$a_{2n-2}$} circle (2pt); 
	\filldraw (2.5,0) node[below] {$a_{2n-1}$} circle (2pt); 
	\draw (1.5,0) arc (180:0:.5);
\end{tikzpicture}                                                               
\Bigg\rangle 
\right)
+2.
\eeA
%%%%%%%%%%%%%%%%%%%%%%%%%%%%%%%%%%%
\subsubsection{Inequality for $T_0$:}
%%%%%%%%%%%%%%%%%%%%%%%%%%%%%%%%%%%
$T_{0,j}$ neither creates nor destroys any $\eta$-singlet pairs. Instead, for a
given singlet pair, it moves one of the doubly-occupied/empty sites in the pair
either left or right one site i.e. it extends the bond distance of the pair by
$\pm1$. The case where the distance \emph{increases} i.e. saturates the
inequality, can be represented
graphically as
\beA
\mathcal{D}\left( T_{0,2n} 
\Bigg\vert                                                                      
\begin{tikzpicture}[scale=0.5, every node/.style={scale=0.55},baseline=-.2ex]                                   
	\filldraw (1.5,0) node[below] {$a_{2n-1}$} circle (2pt); 
	\filldraw (2.7,0) node[below] {$a_{2n}$} circle (2pt); 
	\draw (1.5,0) arc (180:0:.6);
\end{tikzpicture}                                                               
\Bigg\rangle 
\right)
&
\leq
\mathcal{D}\left( 
\Bigg\vert                                                                      
\begin{tikzpicture}[scale=0.5, every node/.style={scale=0.55},baseline=-.2ex]                                   
	\filldraw (1.5,0) node[below] {$a_{2n-1}$} circle (2pt); 
	\filldraw (2.7,0) node[below] {$a_{2n}+1$} circle (2pt); 
	\draw[fill=white,densely dotted] (2.4,0) circle (2pt); 
	\draw (1.5,0) arc (180:0:.6);
	\draw[->] (2.4,-0.1) arc (-150:-30:.15);
\end{tikzpicture}                                                               
\Bigg\rangle 
\right),\\
&
\leq
\mathcal{D}\left( 
\Bigg\vert                                                                      
\begin{tikzpicture}[scale=0.5, every node/.style={scale=0.55},baseline=-.2ex]                                   
	\filldraw (1.5,0) node[below] {$a_{2n-1}$} circle (2pt); 
	\filldraw (2.7,0) node[below] {$a_{2n}$} circle (2pt); 
	\draw (1.5,0) arc (180:0:.6);
\end{tikzpicture}                                                               
\Bigg\rangle 
\right)+1.
\eeA
%%%%%%%%%%%%%%%%%%%%%%%%%%%%%%%%%%%
\subsubsection{Inequality for $T_{-1}$:}
%%%%%%%%%%%%%%%%%%%%%%%%%%%%%%%%%%%
It is simple to show that $T_{-1,j}$ ``fuses'' the ends of singlet pairs
together to create a new singlet pair between the unaffected sites.
Using this, we can consider the two possible distinct cases graphically.
Explicit calculation shows that this intuition holds. The cases are:
(i) $a_{2n-1} = a_{2n}-1$ i.e. they are adjacent
\beA
&
\mathcal{D}\left( 
T_{-1,a_{2n-1}}
\Bigg\vert                                                                      
\begin{tikzpicture}[scale=0.5, every node/.style={scale=0.55},baseline=-.2ex]                                   
	\filldraw (1.5,0) node[below] {$a_{2n-2}$} circle (2pt); 
	\filldraw (2.5,0) node[below] {$a_{2n-1}$} circle (2pt); 
	\filldraw (3.5,0) node[below] {$a_{2n}$} circle (2pt); 
	\filldraw (4.5,0) node[below] {$a_{2n+1}$} circle (2pt); 
	\draw (1.5,0) arc (180:0:.5);
	\draw (3.5,0) arc (180:0:.5);
\end{tikzpicture}                                                               
\Bigg\rangle 
\right)\\
&
\leq
\mathcal{D}\left(
\Bigg\vert                                                                      
\begin{tikzpicture}[scale=0.5, every node/.style={scale=0.55},baseline=-.2ex]                                   
	\filldraw (1.5,0) node[below] {$a_{2n-2}$} circle (2pt); 
	\filldraw (4.5,0) node[below] {$a_{2n+1}$} circle (2pt); 
	\draw[densely dotted] (2.5,0) circle (2pt); 
	\draw[densely dotted] (3.5,0)  circle (2pt); 
	\draw[<->] (2.5,-0.1) arc (-150:-30:.55);
	\draw (1.5,0) arc (135:45:2.1213);
\end{tikzpicture}                                                               
\Bigg\rangle 
\right),\\
&
\leq
\mathcal{D}\left(
\Bigg\vert                                                                      
\begin{tikzpicture}[scale=0.5, every node/.style={scale=0.55},baseline=-.2ex]                                   
	\filldraw (1.5,0) node[below] {$a_{2n-2}$} circle (2pt); 
	\filldraw (2.5,0) node[below] {$a_{2n-1}$} circle (2pt); 
	\filldraw (3.5,0) node[below] {$a_{2n}$} circle (2pt); 
	\filldraw (4.5,0) node[below] {$a_{2n+1}$} circle (2pt); 
	\draw (1.5,0) arc (180:0:.5);
	\draw (3.5,0) arc (180:0:.5);
\end{tikzpicture}                                                               
\Bigg\rangle 
\right),
\eeA
(ii) $a_{2n} = a_{2n+1}-1$ i.e. two singlets are nested
\beA
&
\mathcal{D}\left( 
T_{-1,a_{2n}}
\Bigg\vert                                                                      
\begin{tikzpicture}[scale=0.5, every node/.style={scale=0.55},baseline=-.2ex]                                   
	\filldraw (1.5,0) node[below] {$a_{2n-2}$} circle (2pt); 
	\filldraw (2.5,0) node[below] {$a_{2n-1}$} circle (2pt); 
	\filldraw (3.5,0) node[below] {$a_{2n}$} circle (2pt); 
	\filldraw (4.5,0) node[below] {$a_{2n+1}$} circle (2pt); 
	\draw (2.5,0) arc (180:0:.5);
	\draw (1.5,0) arc (135:45:2.1213);
\end{tikzpicture}                                                               
\Bigg\rangle 
\right)\\
&
\leq
\mathcal{D}\left(
\Bigg\vert                                                                      
\begin{tikzpicture}[scale=0.5, every node/.style={scale=0.55},baseline=-.2ex]                                   
	\filldraw (1.5,0) node[below] {$a_{2n-2}$} circle (2pt); 
	\filldraw (2.5,0) node[below] {$a_{2n-1}$} circle (2pt); 
	\draw[densely dotted] (3.5,0)  circle (2pt); 
	\draw[densely dotted] (4.5,0)  circle (2pt); 
	\draw (1.5,0) arc (180:0:.5);
	\draw[<->] (3.5,-0.1) arc (-150:-30:.55);
\end{tikzpicture}                                                               
\Bigg\rangle 
\right),\\
&
\leq
\mathcal{D}\left( 
\Bigg\vert                                                                      
\begin{tikzpicture}[scale=0.5, every node/.style={scale=0.55},baseline=-.2ex]                                   
	\filldraw (1.5,0) node[below] {$a_{2n-2}$} circle (2pt); 
	\filldraw (2.5,0) node[below] {$a_{2n-1}$} circle (2pt); 
	\filldraw (3.5,0) node[below] {$a_{2n}$} circle (2pt); 
	\filldraw (4.5,0) node[below] {$a_{2n+1}$} circle (2pt); 
	\draw (2.5,0) arc (180:0:.5);
	\draw (1.5,0) arc (135:45:2.1213);
\end{tikzpicture}                                                               
\Bigg\rangle 
\right).
\eeA

%%%%%%%%%%%%%%%%%%%%%%%%%%%%%%
\subsection{Higher Dimensions}
%%%%%%%%%%%%%%%%%%%%%%%%%%%%%%
Any Heisenberg state can be written in the form \fr{Heisstate2D}
\be
|\psi\rangle=\sum_{|\mathfrak{C};\boldsymbol{\sigma}\rangle\in\mathfrak{B}}
A_{\mathfrak{C};\boldsymbol{\sigma}}\
|\mathfrak{C};\boldsymbol{\sigma}\rangle\ ,
\ee
so that in order to establish \fr{ineq} it suffices to show that
\be
\mathcal{D}\left( T_n |\mathfrak{C};\boldsymbol{\sigma}\rangle \right) \leq
\mathcal{D}\left( |\mathfrak{C};\boldsymbol{\sigma}\rangle\right) + n + 1\ .
\ee
for all basis states $|\mathfrak{C};\boldsymbol{\sigma}\rangle\in\mathfrak{B}$.
We consider the three cases $n=0,\pm 1$ in turn.
%%%%%%%%%%%%%%%%%%%%%%%%%%%%%%%%%%%
\subsubsection{Inequality for $T_0$:}
%%%%%%%%%%%%%%%%%%%%%%%%%%%%%%%%%%%
We note that by construction we have for all basis states
\be
\mathcal{D}(|\mathfrak{C};{\boldsymbol\sigma}\rangle) =
\overline{\mathcal{D}}(|\mathfrak{C};{\boldsymbol\sigma}\rangle)\ ,\
|\mathfrak{C};{\boldsymbol\sigma}\rangle\in\mathfrak{B}.
\label{eq:Minequality}
\ee
Following through the same steps as in the one dimensional case we see that
$T_0$ can increase the Manhattan distance between any pair in the
configuration by at most 1 i.e. 
\beA
\mathcal{D}\left( T_0|\mathfrak{C};{\boldsymbol\sigma}\rangle \right)
&
\leq 
\overline{\mathcal{D}}\left( T_0|\mathfrak{C};{\boldsymbol\sigma}\rangle \right),
\\
&
\leq \overline{\mathcal{D}}\left(|\mathfrak{C};{\boldsymbol\sigma}\rangle
\right)+1=\mathcal{D}(|\mathfrak{C};{\boldsymbol\sigma}\rangle),
\eeA
where in the last step we used \fr{eq:Minequality}.
%%%%%%%%%%%%%%%%%%%%%%%%%%%%%%%%%%%
\subsubsection{Inequality for $T_1$:}
%%%%%%%%%%%%%%%%%%%%%%%%%%%%%%%%%%%
Similarly, acting with $T_1$ creates a new configuration with an additional
edge
\beA
\mathcal{D}\left( T_1|\mathfrak{C};{\boldsymbol\sigma}\rangle \right)
&
\leq 
\overline{\mathcal{D}}\left( T_1|\mathfrak{C};{\boldsymbol\sigma}\rangle \right),
\\
&
\leq 
\overline{\mathcal{D}}\left(|\mathfrak{C};{\boldsymbol\sigma}\rangle
\right)+2
=\mathcal{D}\left( |\mathfrak{C};{\boldsymbol\sigma}\rangle \right)+2, 
\eeA
where in the last step we used \fr{eq:Minequality}.
%%%%%%%%%%%%%%%%%%%%%%%%%%%%%%%%%%%
\subsubsection{Inequality for $T_{-1}$:}
%%%%%%%%%%%%%%%%%%%%%%%%%%%%%%%%%%%
Finally, acting with $T_{-1}$ always removes a pair of ``adjacent''
unoccupied/doubly occupied sites. Thinking in terms of configurations, 
this then either removes a paired bond, or fuses two bonds into one.
If we explicitly write the points defining the configuration $\mathfrak{C}$,
the first case corresponds to
\beA
T_{-1}&:
\left\{
	\{ k_1,k_2\},\{k_3,k_4\},\dots,\{k_{2q-1},k_{2q}\}
\right\}\\
&
\to
\left\{
	\{k_3,k_4\},\dots,\{k_{2q-1},k_{2q}\}
\right\}.
\eeA
This implies that
\be
\overline{\mathcal{D}}\left(T_{-1} |\mathfrak{C};\boldsymbol\sigma\rangle \right)
\leq \overline{\mathcal{D}}\left( |\mathfrak{C};\boldsymbol\sigma\rangle \right)-
\left( \lVert k_1 - k_2 \rVert +1 \right).
\ee
The second case corresponds to
\beA
T_{-1}&:
\left\{
	\{ k_1,k_2\},\{k_3,k_4\},\dots,\{k_{2q-1},k_{2q}\}
\right\}\\
&
\to
\left\{
	\{k_1,k_4\},\dots,\{k_{2q-1},k_{2q}\}
\right\},
\eeA
which implies that
\beA
\overline{\mathcal{D}}\left(T_{-1} |\mathfrak{C};\boldsymbol\sigma\rangle
\right)&\leq \overline{\mathcal{D}}\left( |\mathfrak{C};\boldsymbol\sigma\rangle
\right)+ \left( \lVert k_1 - k_4 \rVert +1 \right)\\
&- \left( \lVert k_1 - k_2 \rVert +1 \right) 
- \left( \lVert k_3 - k_4 \rVert +1 \right).
\eeA
By construction $k_2$ and $k_3$ are ``adjacent'', so that
\beA
& \left( \lVert k_1 - k_4 \rVert +1 \right)
- \left( \lVert k_1 - k_2 \rVert +1 \right)\\
&
- \left( \lVert k_3 - k_4 \rVert +1 \right)
\leq 0.
\eeA
Putting everything together we have
\beA
{\mathcal{D}}\left(T_{-1} |\mathfrak{C};\boldsymbol\sigma\rangle \right)
\leq
{\mathcal{D}}\left( |\mathfrak{C};\boldsymbol\sigma\rangle \right),
\qquad |\mathfrak{C};\boldsymbol\sigma\rangle \in \mathcal{B}.
\eeA


\begin{thebibliography}{99}

\bibitem{GM:col_rev02} 
M. Greiner, O. Mandel, T.W. H\"ansch and I. Bloch, 
Nature {\bf 419}, 51-54 (2002). 

\bibitem{kww-06}
T. Kinoshita, T. Wenger and D. S. Weiss,  Nature {\bf 440}, 900 (2006).

\bibitem{HL:Bose07} 
S. Hofferberth, I. Lesanovsky, B. Fischer, T. Schumm and J. Schmiedmayer, 
Nature {\bf 449}, 324-327 (2007). 

\bibitem{hacker10}
L. Hackermuller, U. Schneider, M. Moreno-Cardoner, T. Kitagawa,
S. Will, T. Best, E. Demler, E. Altman, I. Bloch and B. Paredes,
Science {\bf 327}, 1621 (2010).

\bibitem{tetal-11}
S. Trotzky Y.-A. Chen, A. Flesch, I. P. McCulloch, U. Schollw\"ock,
J. Eisert and I. Bloch, Nature Phys. {\bf 8}, 325 (2012). 

\bibitem{getal-11}
M. Gring, M. Kuhnert, T. Langen, T. Kitagawa, B. Rauer, M. Schreitl, I. Mazets, D. A. Smith, E. Demler and J. Schmiedmayer,
Science {\bf 337}, 1318 (2012).

\bibitem{cetal-12}
M. Cheneau, P. Barmettler, D. Poletti, M. Endres, P. Schauss,
T. Fukuhara, C. Gross, I. Bloch, C. Kollath and S. Kuhr,
Nature {\bf 481}, 484 (2012).

\bibitem{langen13}
T. Langen, R. Geiger, M. Kuhnert, B. Rauer, and J. Schmiedmayer, 
Nature Physics {\bf 9}, 640 (2013).

\bibitem{MM:Ising13} 
F. Meinert, M.J. Mark, E. Kirilov, K. Lauber, P. Weinmann, A.J. Daley,
and H.-C. N\"agerl, Phys. Rev. Lett. {\bf 111}, 053003 (2013).  

\bibitem{zoran1}
N. Navon, A.L. Gaunt, R.P. Smith and Z. Hadzibabic,
Science {\bf 347}, 167 (2015).
 
\bibitem{MBLex1}
M. Schreiber, S.S. Hodgman, P. Bordia, H.P. L\"uschen, M.H. Fischer, 
R. Vosk, E. Altman, U. Schneider and I. Bloch, Science {\bf 349}, 842 (2015).

\bibitem{Deutsch91}
J.~M.~Deutsch, Phys. Rev. A {\bf 43}, 2046 (1991).

\bibitem{Sred1}
M.~Srednicki, Phys. Rev. E {\bf 50}, 888 (1994).

\bibitem{Sred2}
M.~Srednicki, J. Phys. A {\bf 32}, 1163 (1998).

\bibitem{ETH}
L. D'Alessio, Y. Kafri, A. Polkovnikov and M. Rigol,
Adv. Phys. 65, 239 (2016).

\bibitem{GGE}
M. Rigol, V. Dunjko, V. Yurovsky and M. Olshanii,
Phys. Rev. Lett. {\bf 98}, 50405 (2007). 

\bibitem{EFreview}
F.~H.~L.~Essler and M.~Fagotti J. Stat. Mech. 064002 (2016).  

\bibitem{Cazalilla06}
M. A. Cazalilla, Phys. Rev. Lett. 97, 156403 (2006).

\bibitem{cc-07}
P. Calabrese and J. Cardy, J. Stat. Mech. P06008, (2007).

\bibitem{CEF}
P. Calabrese, F.H.L. Essler and M. Fagotti,
Phys. Rev. Lett. {\bf 106}, 227203 (2011).

\bibitem{GGE_int} 
E. Ilievski, J. De Nardis, B. Wouters, J.-S. Caux,
F.H.L. Essler and T. Prosen, Phys. Rev. Lett. {\bf 115}, 157201 (2015). 

\bibitem{MBL1}
I. V. Gornyi, A. D. Mirlin and D. G. Polyakov, Phys.
Rev. Lett. 95, 206603 (2005).

\bibitem{MBL2}
D. M. Basko, I. L. Aleiner and B. L. Altshuler, Annals of
Physics {\bf 321}, 1126 (2006).

\bibitem{MBL3}
V. Oganesyan and D. A. Huse, Phys. Rev. B{\bf 75}, 155111 (2007).

\bibitem{MBL4}
A. Pal and D. A. Huse, Phys. Rev. B{\bf 82}, 174411 (2010).

\bibitem{MBL5}
B. Bauer and C. Nayak, J. Stat. Mech. 09005 (2013).

\bibitem{MBL6}
J. Z. Imbrie, J. Stat. Phys. {\bf 163}, 998 (2016).

\bibitem{MBL7}
M. Serbyn, Z. Papi´c, and D. A. Abanin, Phys. Rev. Lett.
111, 127201 (2013).

\bibitem{MBL8}
D. A. Huse, R. Nandkishore, and V. Oganesyan, Phys.
Rev. B{\bf 90}, 174202 (2014).

\bibitem{qdl}
T.~Grover and M.~P.~A.~Fisher, J. Stat. Mech. P10010 (2014).

\bibitem{hubb1}
J.R. Garrison, R.V. Mishmash and M.P.A. Fisher,
arXiv:1606.05650v1

\bibitem{takahashi}
M. Takahashi, J. Phys. {\bf C 10}, 1289 (1977).

\bibitem{MacDonaldGirvin88}
A.~H.~MacDonald, S.~M.~Girvin, D.~Yoshioka, Phys. Rev. B {\bf 37}, 9753 (1988).

\bibitem{stein}
J.~Stein, J. Stat. Phys. {\bf 88} 487 (1997).

\bibitem{sasha}
A. L. Chernyshev, D. Galanakis, P. Phillips, A. V. Rozhkov and
A.-M. S. Tremblay, Phys. Rev. B{\bf 70}, 235111 (2004).

\bibitem{etapairing}
O.~J. Heilmann and E.~H. Lieb, Ann.~N.Y.~Acad.~Sci. \textbf{172}, 584
(1971); C. N. Yang, Phys. Rev. Lett. {\bf 63}, 2144 (1989).

\bibitem{book}         
F.~H.~L.~Essler, H.~Frahm, F.~G{\"o}hmann, A.~Kl{\"u}mper, and
V.~E.~Korepin, {\it The One-Dimensional Hubbard Model},
Cambridge University Press, Cambridge (2005).

\bibitem{CE13}
J.S. Caux and F.H.L. Essler, Phys. Rev. Lett. {\bf 110}, 257203 (2013).

\bibitem{takahashiTBA1}
M. Takahashi, Prog. Theor. Phys. {\bf 47}, 69 (1972).

\bibitem{takahashiTBA2}
M. Takahashi, Prog. Theor. Phys. {\bf 52}, 103 (1974).

\bibitem{Ha}
Z.N.C. Ha, Phys. Rev. B{\bf 46}, 12205 (1992).

\bibitem{EEG}
S. Ejima, F.H.L. Essler and F. Gebhard, J. Phys. {\bf A39}, 4845 (2006).
\bibitem{temperleyLieb}
		H.~N.~V.~Temperley and E.~H.~Lieb, Proc. R. Soc. London A {\bf
		322}, 251 (1971).
\bibitem{saitoproof}
		R.~Saito, J. Phys. Soc. Jpn. {\bf 59}, 482 (1990).

\bibitem{takahashigsE}
M. Takahashi, Prog. Theor. Phys. {\bf 45}, 756 (1971).

\end{thebibliography}
\end{document}